\documentclass[12pt]{iopart}
\usepackage{epsfig}
\usepackage{graphicx}
\usepackage{lscape}
\usepackage{longtable}

\begin{document}

\title[$\alpha$ Persei Debris Disks]{Stellar Membership and Dusty Debris Disks in the $\alpha$ Persei Cluster}

\author{B. Zuckerman$^1$,Carl Melis$^2$,Joseph H. Rhee$^3$,Adam Schneider$^4$,Inseok Song$^4$}

\address{$^1$Department of Physics and Astronomy, University of California, Los Angeles, CA 90095, USA}
\address{$^2$Center for Astrophysics and Space Sciences, University of California, San Diego, CA 92093, USA}
\address{$^3$Department of Physics and Astronomy, California State Polytechnic University, Pomona,
3801 W. Temple Ave., Pomona, CA 91768, USA}
\address{$^4$Department of Physics and Astronomy, University of Georgia, Athens, GA 30602-2451, USA} 

\begin{abstract}
Because of proximity to the Galactic plane, reliable identification of members of the $\alpha$ Persei cluster is often problematic.  Based primarily on membership evaluations contained in six published papers, we constructed a mostly complete list of high-fidelity members of spectral type G and earlier that lie within 3 arc degrees of the cluster center.   $\alpha$ Persei was the one nearby, rich, young open cluster not surveyed with the Spitzer Space Telescope.  We examined the first and final data releases of the Wide Field Infrared Survey Explorer (WISE) and found 11, or perhaps 12, $\alpha$ Per cluster members that have excess mid-infrared emission above the stellar photosphere attributable to an orbiting dusty debris disk.  The most unusual of these is V488 Per, a K-type star with an excess IR luminosity 16\% (or more) of the stellar luminosity; this is a larger excess fraction than that of any other known dusty main sequence star.  Much of the dust that orbits V488 Per is at a temperature of $\sim$800 K; if these grains radiate like blackbodies, then they lie only $\sim$0.06 AU from the star.  The dust is probably the aftermath of a collision of two planetary embryos or planets with small semimajor axes; such orbital radii are similar to those of many of the transiting planets discovered by the Kepler satellite. 
\end{abstract}
\maketitle

\section{INTRODUCTION}

Unveiling the evolution of dusty debris disks as a function of stellar age was a major focus of studies of main sequence stars with the Spitzer Space Telescope.  Rebull et al (2008) and Gaspar et al (2009) each presented tables listing stellar associations and clusters of known age and the fraction of these stars with Spitzer-detected excess infrared emission.  In a recent paper, we added the AB Doradus, Tucana/Horologium, and Argus associations to such tabulations (Zuckerman et al 2011, Table 11).

The $\alpha$ Persei cluster (hereafter $\alpha$ Per) is the only nearby (182$\pm$11 pc, Jackson \& Jefferies 2010; 191$\pm$7 pc, Robinchon et al. 1999; 183 pc, Makarov 2006), rich (hundreds of members), open cluster not surveyed by Spitzer; indeed, at an age of $\sim$60 Myr (Section 3.2) it is the youngest rich open cluster in the solar vicinity.  To partially rectify this omission we correlated mid-infrared sources in the first and final WISE data releases with members of $\alpha$ Per that have spectral classes between M- and B-type, but with most emphasis on G-type and earlier.  Because one of our goals was determination of the fraction of cluster members with WISE-detected mid-IR emission above the stellar photosphere, it was important to identify stars with a high probability of membership.  Toward this end, we evaluated results from various published papers to compose lists of likely and possible cluster members (Tables 1 and 2, respectively).   The list in Table 1 should be mostly complete for stars of spectral type G and earlier and located within 3 degrees of the $\alpha$ Persei cluster center which we took to be 03h26.0m +49d07' (Kharchenko et al 2005).    

In the following sections we outline our search procedure for dusty $\alpha$ Per stars, discuss a few of the dusty stars that appear in Table 3, and consider where $\alpha$ Per stands relative to other young open clusters and associations as regards the presence of dusty debris disks.  In compiling Tables 1 and 2 we had occasion to consult many of the important papers written about the $\alpha$ Persei cluster.   These and additional important papers are cited in Section 3.2 where we consider the relative ages of $\alpha$ Persei and other nearby stellar clusters and associations.

\section{SEARCH PROCEDURE FOR EXCESS INFRARED EMISSION}

Our identification of $\alpha$ Per members with dusty debris disks involved
three principal steps.  First we produced a list of likely cluster members.
We correlated this list with the WISE catalog, and then searched for stars with infrared emission above the stellar photosphere.

\subsection{$\alpha$ Persei Input Samples}

To maintain as wide a breadth as possible in the early stages of our search,
we employed three cluster input catalogs. The first was taken from
the WEBDA online service (http://www.univie.ac.at/webda/), another was 
from Makarov (2006), and the third was generated by ourselves from a focused search of the
UCAC3 catalog (Zacharias et al 2009). All sources listed in the WEBDA $\alpha$
Persei catalog (regardless of membership probability) were searched for
infrared excess emission (see Section 2.2), as were all 139 stars
listed in Table 1 of Makarov (2006) and identified by him as "high fidelity" members.  Because 
Makarov's list is mostly complete only to spectral type mid-G and earlier, our search of the
UCAC3 catalog was designed to reveal late-type stars with exceptionally large
mid-IR excess emission similar to and in addition to that of V488 Per (Section 3.1, Figure 2), but none were found.

The UCAC3 input sample was generated as follows. We searched the UCAC3 catalog
within 3 arc degrees of the $\alpha$ Persei cluster center at 03h26.0m +49d07'.
To restrict the size of the search output, we
further required that a star have proper motion errors of less than 10
mas/yr in each of RA and DEC, that pmRA lie between 7 and 37 mas/yr, and
that pmDEC lie between -40 and -10 mas/yr. The proper motion of stars
meeting these criteria were compared to the mean $\alpha$ Persei proper motion
values (pmRA=21.98$\pm$0.20 mas/yr, pmDEC=-25.36$\pm$0.17 mas/yr; Kharchenko et
al. 2005). If the UCAC3 proper motions of a queried source agreed with the
mean cluster proper motion values to within 3-sigma of their respective
uncertainties, then the source was tentatively labeled as a cluster member.
All tentative cluster members were then searched for infrared excess emission.

The WEBDA service collects data products from papers that discuss open
clusters. The service makes no effort to vet these data products, and merely
adds them as-is to their repository. This means that if, for example, a paper
were to report imaging photometry of sources in the field around $\alpha$
Persei, all such sources would be added to the WEBDA $\alpha$ Persei cluster
database regardless of actual membership status. In this way the WEBDA
service collected
$>$3000 candidate $\alpha$ Persei cluster members. To be as conservative as 
possible, and to enable as much infrared discovery as possible, we started
with all 3000+ candidates given by WEDBA for the $\alpha$ Persei cluster. These
candidates come from 14 sources, three of which are cited in the present paper in other contexts.

To compile excess statistics, cluster members both with and without infrared
excess emission needed to be culled from our larger input catalogs. Proper
motion information, photometric distance estimates, and radial velocity
measurements (when available) were used to identify members and to remove
non-members from our input samples. In this way the majority of the WEBDA
input sample was rejected.  

Following our own search, as outlined in the preceding four paragraphs, we then turned
to the $\alpha$ Persei cluster literature.  In addition to Makarov (2006), we principally 
utilized six other papers (Tables 1 and 2, see the six right hand columns and the table
notes) to identify likely and possible cluster members.  In addition to proper motion, Hipparcos
and photometric distances, and radial velocities, lithium abundance and X-ray fluxes have
been utilized to establish cluster membership in these various papers.

Makarov's (2006) selection criteria are described in the first two sentences
of his abstract: "A kinematical study of the nearby open cluster $\alpha$ Persei
is presented based on astrometric proper motions and positions in the Tycho-2
catalog and Second USNO CCD Astrographic Catalog (UCAC2). Using the
astrometric data and photometry from the Tycho-2 and ground-based catalogs,
139 probable members of the cluster are selected, 18 of them new."  

As noted
two paragraphs previous, we and others have used criteria in addition to
those employed by Makarov.  The result is that 117 of Makarov's 139 suggested
members appear in our Table 1 and 20 appear in our Table 2.
As may be seen by inspection of Table 2, the principal reasons why 20 stars
from Makarov's list are deemed by us to be only possible cluster members is
because either other authors do not agree with Makarov or because no data
other than kinematics and photometry are available.  By contrast, almost all
stars listed in Table 1 are regarded as members in at least two papers.  In
the few cases where an "n" (nonmember) or "n?" appears in Table 1, such
characterization is often superceded in later references, sometimes by the
same authors.

While Table 1 should be a mostly complete list of members of spectral type G and 
earlier and located within 3 degrees of the cluster center, the table only sparsely samples later-type
members (e.g., AP members listed by Prosser 1992).  Because of WISE sensitivity limits that would 
require huge fractional excess IR emission for cluster members later than spectral type G, we have 
not made an effort to identify such stars.   Therefore, many previously suggested late-type cluster 
members do not appear in Tables 1 and 2.    Also, although we considered X-ray 
data when compiling Tables 1 and 2 (e.g. Randich et al 1996), we did not do a thorough search of
all published X-ray papers that might be relevant to identification of cluster members.

\subsection{WISE Cross-correlation}

Each input catalog of $\alpha$ Persei stars was initially cross-correlated with the preliminary (2011) WISE
catalog.  Given the WISE angular resolution of 6.1", 6.4", 6.5" and 12.0" in its four bands of 3.35, 4.60, 11.56, and 22.09 $\mu$m (Wright et al 2010), we used a search radius of 10" to identify WISE counterparts of input catalog stars.  For conversion from WISE magnitudes to the flux densities given in Table 3 we used 0 (Vega-magnitude) flux densities of 306.68, 170.66, 29.045, and 8.284 Jy and color corrected these as appropriate for each star.

Spectral energy distributions (SEDs) for each matched source were generated using 2MASS photometry, WISE data, and any other
available photometry (e.g., Tycho-2 BV magnitudes).  A fully automated SED-fitting technique employing a theoretical atmospheric model (Hauschildt et al 1999) was used to predict stellar photospheric fluxes.  A detailed description of our photospheric fitting procedure is given in Section 2 of Rhee et al (2007), so we do not repeat it here.  In Section 3 of our recent paper (Zuckerman et al 2011) we describe how we checked this theoretical model against actual MIPS measurements of stars that show no evidence for an infrared excess.  The result was an upward correction of the model photospheric flux densities at 24 um by a factor of 1.03.  Therefore, in Table 3 the listed 22 $\mu$m photospheric flux densities are those produced from the theoretical model, multiplied by a factor of 1.03.

From a sample of 149 likely cluster members (Table 1), mostly B-, A-, F- and G-type stars, in the 2011 WISE catalog we initially identified a total of 11 $\alpha$ Per cluster members with excess mid-IR emission, probably due to orbiting dust particles (Table 3; Figures 1-2).   The final (2012) version of the WISE catalog appeared shortly before this paper was finalized and the same 11 stars were identified; all Table 3 flux densities are from the final WISE catalog.  These 11 include V488 Per which is the only star with excess emission in each of the four WISE filters (see discussion of V488 Per in Section 3.1).  For each of the 11 stars in Table 3, the 22 $\mu$m excess is at least 5 times the listed WISE uncertainty.  In addition to these 11 stars, HD 21855 might have a 22 $\mu$m excess (see discussion in Section 3.1).

\subsection{Spitzer/IRAC Observations of V 488 Per}

V488 Per was serendipitously observed with the Infrared Array Camera (IRAC;
Fazio et al. 2004) on Spitzer through a program searching for substellar
companions to low mass stars in $\alpha$ Per (AORKEY: 18853632). It was
observed in all four IRAC channels on 9-26-2006.  IRAC Post Basic
Calibrated Data (pBCD) mosaic files were obtained from the Spitzer
Heritage Archive.  Source extraction was performed on the pBCD mosaics
utilizing the Astronomical Point Source Extraction (APEX) package,  and
the IRAC data points are added to the SED of V488 Per to further
constrain the dust fitting (Figure 2).  The Spitzer IRAC observations
support the large L$_{IR}$/L* indicated by WISE for this object.

\section{DISCUSSION}

We identified 11 stars with excess emission above the stellar photosphere at 22 $\mu$m and sometimes at other wavelengths (Table 3, Figures 1 and 2) that can definitely or probably be attributed to an orbiting dusty debris disk.  While most are of early-type -- for which the detected dust emission is usually the Wien tail of emission at longer wavelengths generated by cool dust particles --  one remarkable K-type star, V488 Per, appears in Tables 1 and 3 and Figure 2.  

The distribution of spectral types in Table 3 is five B, three A, two F and one K.   Some additional B-type stars display excess emission as measured by WISE, but these excesses are more likely due to free-free radiation in an outflowing ionized stellar wind rather than from an orbiting debris disk.   Characteristically, when emission is due to free-free transitions, then IR flux densities decline toward longer wavelengths.  One prominent Table 1 example of such a star is $\psi$ Per, strongly detected with IRAS many years ago.   Another such B-type star is HR 1037 (= HD 21362; compare the 24 and 70 $\mu$m MIPS measurements reported by Su et al (2006) with the WISE catalog flux densities).  

Below we consider a few stars in Table 3 and also the stars 29 Per and HD 21855.  These are followed by a discussion of the evolution of dusty debris disks.

\subsection{Individual SEDs}

29 Per:  This B3 star, whose SED is shown in Figure 1, might be expected to display an outflowing ionized wind as mentioned just above.  However, the shape of the SED of 29 Per is very different from that of other B-type stars.  The excess emission at 22 $\mu$m is weak, whereas IRAS measured strong excess emission at 60 $\mu$m (Figure 1).  We therefore inspected the WISE 22 $\mu$m images and also used the IRAS "scanpi" function to characterize the spatial distribution of the 60 $\mu$m emission in the vicinity of 29 Per.  The WISE image displayed in Figure 3 illustrates the extended nature of the 22 $\mu$m emission.  Our conclusion is that the excess emission is carried by heated interstellar dust ("cirrus") at substantial distances from 29 Per and not to an orbiting debris disk.  Thus, 29 Per is included in Figure 1 to illustrate its SED, but not to suggest the presence of a debris disk.

\vskip 0.1in				
\noindent V488 Per:  This solar-type star has a larger fractional infrared luminosity, defined as the ratio of excess infrared luminosity to bolometric luminosity ($\tau$), than any other known main sequence star.   From the SED in Figure 2, $\tau$ $\sim$16\%, to be compared to the previous main sequence champion BD+20 307 with $\tau$ $\sim$3.3\%.  (Song et al 2005; Weinberger et al 2011).   Most main sequence stars with strong mid-IR emission from warm dust particles show no evidence for substantial quantities of cool dust (e.g.. Melis et al 2010; Melis et al 2012; Weinberger et al 2011; and references therein).  However, as may be seen from blackbody fits to its SED, substantial quantities of both warm and cool dust particles orbit V488 Per.

Much of the dust that orbits V488 Per is at a temperature of $\sim$800 K.  While comparably warm dust with $\tau$ orders of magnitude smaller than at V488 Per may orbit some main sequence stars, such small quantities of dust could easily be generated by asteroidal collisions near these stars.  By contrast, the large $\tau$ at V488 Per indicates a recent collision of planet-mass objects (Melis et al 2010).   If these warm grains radiate like blackbodies, then they lie only $\sim$0.06 AU from the star.  This semimajor axis is comparable to that characteristic of many transiting planets discovered by NASA's Kepler satellite. 

Based on the data given in various published papers -- these include photometric distance, proper motion, radial velocity, lithium abundance, and chromospheric activity -- there seems to be little doubt that V488 Per is a bona fide member of the $\alpha$ Per cluster and it has been so classified in three of the papers cited in Table 1.  In addition we can add new determinations of radial velocity, proper motion, and lithium abundance that buttress previous work on this star.

Concerning radial velocity, archival Keck/HIRES observations of V488 Per obtained by Dr. I. N. Reid on UT 12 December 1997 were acquired from the Keck Observatory Archive\footnote{http://www2.keck.hawaii.edu/koa/public/koa.php}.  
We analyzed this high quality echelle spectrum and found a heliocentric velocity of 1.8$\pm$0.5 km/s which agrees reasonably well with previous radial velocity determinations by Mermilliod et al (2008; -0.31$\pm$0.15 km/s) and Prosser (1992; -0.1$\pm$0.6 km/s).  All of these measured radial velocities are characteristic of stars in $\alpha$ Per (see, e.g., Table 4 in Prosser 1992).

From the HIRES spectrum we measured the equivalent width (EW) of the 6708 \AA\ lithium absorption line to be $\sim$200 m\AA.  This is the EW anticipated for an early K-type star of age $\sim$60 Myr; in Figure 3 of Zuckerman \& Song (2004) this EW lies just below the EW of most of the plotted 30 Myr old members of the Tucana Association and near the top of the range of EW for $\sim$100 Myr old Pleiades members.

We computed the proper motion of V488 Per by comparing positions in the 2MASS and WISE catalogs that are separated by epochs 10.1 years apart.  We derive proper motions of +15.7+/- 6.6 and $-$24.9+/-9.1 mas/yr in R.A. and Decl., respectively.  These values compare well with those given in the USNO-B1 (16, $-$24) and the PPMXL (15.9, $-$26.4) catalogs.

Because $\tau$ is so large for V488 Per compared to all other known dusty main sequence stars, one must consider the possibility of contamination by radiation from a background object.  The best spatial resolution image of the field is likely that obtained with HST by Patience et al (2002).  The HST/NICMOS camera and F140W filter revealed no source of 1.4 $\mu$m emission other than the star.   

To further confirm that the large IR excess is from the star, at our request, Dr. Charles Beichman and collaborators (see caption to Figure 4) obtained a seeing-limited image of V488 Per with a 3.8 $\mu$m (Lp) filter and the NIRC2 camera on the Keck II telescope at Mauna Kea Observatory.  This image shows only a single source of FWHM $\sim$0.6" and approximately circular shape (Figure 4).  Thus, there can be little doubt that the strong IR excess emission is indeed associated with V488 Per.

An additional concern is whether V488 Per might be a dusty, first ascent "Phoenix Giant" star (Melis 2009; Melis et al 2009) well to the background of the $\alpha$ Per cluster.  Some Phoenix Giants have very large $\tau$ and hot and cool dust with SEDs similar to that of V488 Per.  If V488 Per is actually a Phoenix Giant, then the fact that the photometric distance, radial velocity, and proper motion of V488 Per are all consistent with membership in $\alpha$ Per would have to be regarded as simply a strange coincidence.  As outlined in Melis et al (2009), we use line ratio diagnostics presented in Strassmeier \& Fekel (1990) and the Keck/HIRES spectrum to 
evaluate whether V488 Per is a luminosity class III or class V star.  We find that V488 Per is a luminosity class V star.  All K-type Phoenix 
Giants exhibit luminosity class III line ratios when similar analysis is performed for them.

As an aftermath of collisions of planetary embryos, solar-type stars are most likely to display large quantities of warm dust particles during the last stages of the assembly of rocky planets in the terrestrial planet zone; this would be at ages between 30 and 100 Myr (Melis et al 2010) .  The age of V488 Per lies in this range and, so, this very dusty star can be added to the small sample of similar stars discussed by Melis et al.

\vskip 0.1in				
\noindent HD 21375: This is a single-line, 31 day, spectroscopic binary according to Morrell \& Abt (1992).

\vskip 0.1in				
\noindent HD 21641: This is a weak emission, shell, Be star (e.g., Briot 1986).  Therefore we cannot be sure that the observed excess IR emission (Table 3, Figure 2) is due to a dusty debris disk rather than an outflowing ionized wind.  If the latter, then the excess emission becomes apparent only at much longer wavelengths than is seen toward Be stars $\psi$ Per and HR 1037 (both mentioned above).  Future observations at far-IR wavelengths may distinguish between the two potential emission mechanisms. 

\vskip 0.1in				
\noindent HD 21855: This A-type may have weak excess emission at 22 $\mu$m (W4) and perhaps also at 11.5 $\mu$m (W3).   At W4, the photosphere is expected to be 6.4 mJy.  The measured WISE flux density is 10.2$\pm$0.87 mJy, so excess emission at the 4.4 $\sigma$ level may be present.  In the W3 band, the expected photosphere is 22.6 mJy and the WISE measured flux density is 23.5$\pm$0.39 mJy, for a possible 2.3 $\sigma$ excess.

\subsection{Evolution of Dusty Debris Disks with Time}

Table 11 in Zuckerman et al (2011) presented infrared excess fractions at wavelengths of 24 and 70 $\mu$m for various open clusters and nearby moving groups.   Table 4 of the present paper repeats the 24 $\mu$m column from their Table 11, but now with the addition of the 22 $\mu$m excess fraction in the $\alpha$ Per cluster.  Before comparing $\alpha$ Per to other clusters we consider some techniques for estimation of ages of stars in clusters and associations.  


The two principal techniques for derivation of ages of young clusters are
matching cluster members of various spectral types to theoretical pre-main
sequence isochrones and identification of the location of the lithium
depletion boundary.  For a given cluster with age comparable to that of
$\alpha$ Per and for which both age-dating methods have been applied, the
lithium method generally yields an older age. Therefore, it is important to
take into account how ages have been deduced when comparing the age of
$\alpha$ Per to those of other clusters in Table 4.

Figure 5 presents color magnitude diagrams for $\alpha$ Per and for the
Pleiades. The latter is clearly the older cluster.  A quantitative measure
of this age difference is suggested from the Li depletion analysis of
Barrado y Navascues et al (2004) who derive ages of 50$\pm$5, 85$\pm$10, and 130$\pm$20 Myr for IC 2391,
$\alpha$ Per, and the Pleiades, respectively.  In comparison, isochrone ages
for these clusters have been given as 35-40 Myr (IC 2391, Barrado y
Navascues et al 2004; Torres et al 2008), 50-80 Myr ($\alpha$ Per, Makarov
2006; Prosser 1992), and $\sim$100 Myr (Pleiades, see references listed in
Luhman et al 2005).  Discussions of the age of $\alpha$ Per by Makarov
(2006) and by Prosser (1992) illustrate that ages derived from main
sequence fitting can be based on the upper main sequence, the lower main
sequence, or both, and that the former is subject to uncertainties in the
degree of convection overshooting.

The age of the AB Dor moving group has been a subject of considerable
discussion (Luhman et al 2005; Janson et al 2007; Close et al 2007; Ortega
et al 2007) with ages ranging from 50 Myr to the age of the Pleiades ($\leq$120
Myr). This discussion is outlined in Section 8 of Torres et al (2008) who
present their own color-magnitude diagram of AB Dor members and who prefer
an age of 70 Myr.

For M47 Rojo Arellano et al (1997) use main sequence fitting to derive an
age of $\sim$100 Myr, while for NGC 2451 an isochronal age of 60 Myr is deduced
(references listed in Balog et al 2009).

Based on the discussion in the preceding paragraphs, one may see that the
ages in Table 4 are weighted more toward isochrone fitting than Li
depletion.  This is in part because Li ages have been derived for only some
of the listed clusters and associations.  However, whatever the true absolute cluster ages,
the relative ages as given in Table 4 should be quite reliable.

As noted in the Introduction, excepting the $\alpha$ Persei cluster, Spitzer surveyed stars in all the clusters and associations listed in Table 4.  The listed 24 $\mu$m Spitzer/MIPS excess statistics are usually dominated by F, G and K type stars.  In the Zuckerman et al (2011) paper, a ratio of MIPS measured 24 $\mu$m flux density to expected photospheric flux of 1.16 or greater was deemed sufficient to establish the presence of an excess.  This ratio was not uncharacteristic of excess criteria employed in other references listed in Table 4.  Unfortunately, given the size of the WISE catalog errors in flux density at 22 $\mu$m and the greater distance from Earth of the $\alpha$ Per cluster compared to a majority of the regions listed in Table 4, this level of sensitivity to excess emission was obtained only for most of the B-type stars in $\alpha$ Per.  Even for an $\alpha$ Per A-type star, the excess emission would have to be at a level $\sim$60\% of the photosphere for it to satisfy the excess criterion set out in Section 2.2.  

As a result, it is not possible at this time to sensibly compare excess fractions in $\alpha$ Per and other Table 4 clusters.  That is, the excess statistics listed for the $\alpha$ Per cluster pertain only to B- and A-type members and, to qualify as an excess star, the measured 22 $\mu$m flux density must often be as large as 1.5 times the photospheric flux.  In contrast, the statistics for other listed regions in Table 4 usually are dominated by F, G and K-type stars and represent excesses that need only be $\sim$1.2 times the photospheric flux.   Since many stars in, for example, the Zuckerman et al (2011) lists have 24 $\mu$m flux density between 1.2 and 1.5 times that of the photosphere, there can be little doubt that some A-type stars in the $\alpha$ Per cluster have excess emission not far below our defined level of excess and will have been missed with our WISE survey.  Such caveats apply even more strongly to the many F- and G-type stars listed in Tables 1 and 2.

We may summarize what might reasonably be inferred from Table 4 as follows:   Although the $\alpha$ Persei cluster may be deficient in dusty debris disks relative to other Table 4 clusters and moving groups of comparable age, because (1) the $\alpha$ Per statistics pertain only to B- and A-type stars whereas the statistics for the other groups are dominated by later-type stars, and (2) the WISE sensitivity to dusty debris at these $\alpha$ Per stars was not quite as good as Spitzer sensitivities in these other groups, it would be premature to draw any firm conclusions.
Additional infrared measurements at 22 or 24 $\mu$m are unlikely to come before the James Webb Space Telescope is operational.   Before then, perhaps the best way to 
determine if debris disks in $\alpha$ Per are relatively scarce would be to confirm or deny whether the A- and B-type stars in Table 2 are actual cluster members (none have 22 $\mu$m excess emission).  If most are members, then as indicated by statistics given in the Note to Table 4, the implication of a real relative deficiency of debris disks in the $\alpha$ Per cluster would be suggestive, if not totally convincing.

$\alpha$ Per is at the same or smaller absolute Galactic latitude than all Table 4 clusters and moving
groups excepting M47.  Therefore, in the event that background cirrus contamination could introduce an
error in the disk fraction, this error would most likely be in the sense to
increase the apparent number of dusty stars in $\alpha$ Per relative to other clusters and associations in Table 4.  

As remarked in the notes to Table 4, the tabulated excess fraction of the $\alpha$ Per cluster is for a wavelength of 22 $\mu$m, but at 24 $\mu$m for all other entries.  Dusty debris disks studied by IRAS and Spitzer at far-IR wavelengths have typical temperatures of $\sim$80 K.  For this temperature the flux density will be $\sim$1.5 times larger at 24 $\mu$m than at 22 $\mu$m.  Given our 5 $\sigma$ WISE detection threshold at 22 $\mu$m (see Section 2.2), we checked among Table 1 and 2 stars, but not listed in Table 3, for any B- or A-type stars with at least 3.3 $\sigma$ of apparent excess emission; that is, for stars for which "W4 meas" is greater by at least 3.3 $\sigma$ than "W4 Photo" (see the header to Table 3).  Other than HD 21855 with 4.4 $\sigma$ of possible excess emission (see Section 3.1),
no such stars were found.  Thus, the difference between 22 and 24 $\mu$m should not alter the ratio of percentage of excess stars in $\alpha$ Per compared to excess percentages in other listed clusters  

Dukes \& Krumholz (2011; see also references cited therein) suggest that disk disruption events are no more likely in massive clusters than in low mass clusters.  They cite observations that would indicate that, for most clusters, 90\% of the stars disperse into the field within 10 Myr.  
The stellar binary fraction in $\alpha$ Per and other rich nearby clusters (Pleiades, Hyades, Praesepe) have been considered in various papers.  Patience et al (2002) note that the distribution of semimajor axes of binary stars in these four clusters peaks at 5 AU, "a significantly smaller value" than for stars in the solar neighborhood.  Morrell \& Abt (1992) comment on the relatively small number of spectroscopic binaries among B- and A-type stars in $\alpha$ Per and in the Pleiades.   Makarov (2006) notes that counting all types of binaries in $\alpha$ Per, only $\sim$20\% of members are binaries which is "modest even compared with field G-dwarfs and much smaller than the binarity rate in the Hyades and Pleiades."     Calculations -- utilizing the known characteristics of clusters and associations listed in Table 4 -- of the (relative) frequency of close stellar encounters might be of interest.


\section{CONCLUSIONS}

We assembled what likely is the most complete listing of probable and possible B-type through G-type members of the nearby $\alpha$ Persei cluster.  We then correlated sources in the final WISE catalog with cluster members and found 11, or possibly 12, with excess emission above the photosphere at 22 $\mu$m (and sometimes also at other WISE wavelengths), that we interpret as due to a surrounding dusty debris disk.  In addition, a few B-type stars display excess emission consistent with production in an outflowing wind.  Of the probable B-  and A-type cluster members, 14\% display excess emission (from an orbiting dusty debris disk) at a wavelength of 22 $\mu$m.   If B- and A-type stars that are possible members are included in the statistics, the percentage of excess stars could be as small as 11.6\%.  For later-type stars, the WISE sensitivity is inadequate to detect many (most?) of the debris disks that likely orbit stars in the cluster.

V488 Per is the most interesting dusty star in the $\alpha$ Persei cluster.   The luminosity of orbiting dust grains is at least 16\% of the bolometric luminosity of this K-type star; this is a far larger excess luminosity percentage than that previously known at any main sequence star.  Although there are numerous reasons to include V488 Per as a bona fide member of the cluster, for confirmation, a measurement of its trigonometric parallax would be worthwhile.  Recently, Melis et al (2012b) identified a young star with, initially, a fractional excess luminosity at mid-IR wavelengths almost as large as that at V488 Per.  However, the excess at this young star recently disappeared -- in the space of only a few years.  Thus, the excess infrared emission at V488 Per bears careful watching during the coming years.

Previously we considered the final era of the accretion of rocky planets around solar mass stars in a region analogous to that where the four rocky planets of the Solar System reside (Melis et al 2010).   By assembling a list of stars known to contain large quantities of warm dust grains, we found that this era occurs at stellar ages from 30 to 100 Myr.  With its very strong mid-IR excess emission and its $\sim$60 Myr age, V488 Per can now be included among those stars that are orbited by large rocky bodies in the terrestrial planet zone.  By contrast, Melis et al (2012a) find that the dominant era of rocky planet formation around stars with masses a few times that of the Sun is over by about 30 Myr; our study of the $\sim$60 Myr old $\alpha$ Persei cluster is consistent with this finding.  Specifically, among 70 late-B to early-F type $\alpha$ Per cluster members, WISE reveals none with the massive amounts of warm dust that would herald the aftermath of a collision of planetary embryos.

Future observations at far infrared and/or submillimeter will be necessary to characterize the properties of the $\alpha$ Per debris disks.   For V488 Per such observations will be especially important because, unlike most other main sequence stars that are orbited by large quantities of warm dust particles but little or no cold dust, V488 Per appears to be orbited by large amounts of both warm and cold dust. 

\vskip 0.2in

We are grateful to Chas Beichman, Chris Gelino, Greg Mace, Joel Aycock, and Randy Campbell for obtaining the data that went into the image in Figure 4 (just before V488 Per set for the evening), and to Alan Stockton for attempting a similar measurement for us.  We thank John Stauffer for donation of Figure 5 and for helpful conversations and the referee for a constructive report that substantially improved this paper.
This research was funded in part by NASA grants to UCLA and the University of Georgia.  C.M. acknowledges support from the National Science Foundation under award AST-1003318.  This publication makes use of data products from the Wide-field Infrared Survey Explorer, which is a joint project of the University of California, Los Angeles, and the Jet Propulsion Laboratory/California Institute of Technology, funded by the National Aeronautics and Space Administration.  This research has made use of the Keck Observatory Archives (KOA) which is operated by the W.M. Keck Observatory and the NASA Exoplanet Science Institute (NExScI) under contract with the National Aeronautics and Space Administration.

\section*{References}
\begin{harvard}

\item[Balog, Z., Kiss, L., Vinko, J. et al 2009, ApJ 698, 1989]
\item[Barrado y Navascues, D., Stauffer, J. \& Jayawardhana, R. 2004, ApJ 614, 386]
\item[Briot, D. 1986, A\&A 163, 67]
\item[Close, L., Thatte, N., Nielsen, E., et al 2007, ApJ 665, 736]
\item[Dukes, D. \& Krumholz, M. 2011, arXiv1111.3693]
\item[Fazio, G., Hora, J., Allen, L. et al 2004, ApJS 154, 10]
\item[Gaspar, A., Rieke, G. Su, K. et al 2009, ApJ 697, 1578]
\item[Gorlova, N., Rieke, G., Muzerolle, J. et al 2006, ApJ, 649, 1028]
\item[Hauschildt, P., Allard, F. \& Baron, E. 1999, ApJ 512, 377]
\item[Heckmann, V., Dieckvoss, W. \& Kox, H. 1956, AN 283, 109]
\item[Jackson, R. \& Jefferies, R. 2010, MNRAS 402, 1380]
\item[Janson, M., Brandner, W., Lenzen, R. et al 2007, ApJ 462,615]
\item[Kharchenko, N., Piskunov, A., Roser, S., Schilbach, E. \& Scholz, R.-D. 2005, A\&A 438, 1163]
\item[Luhman, K., Stauffer, J. \& Mamajek, E. 2005, ApJ 628, L69]
\item[Makarov, V. 2006, AJ 131, 2967]
\item[Melis, C. 2009, Ph.D. Thesis UCLA]
\item[Melis, C., Zuckerman, B., Rhee, J. \& Song, I. 2010, ApJ 717, L57]
\item[Melis, C., Zuckerman, B., Song, I., Rhee, J., \& Metchev, S. 2009, ApJ 696, 1964]
\item[Melis, C., Zuckerman, B., Rhee, J., et al.  2012a, submitted to ApJ]
\item[Melis, C., Zuckerman, B., Rhee, J., et al.  2012b, submitted to Nature]
\item[Mermilliod, J.-C., Queloz, D. \& Mayor, M. 2008, A\&A 488, 409]
\item[Morrell, N. \& Abt, H. 1992, ApJ 393, 666]
\item[Ortega, V., Jilinski, E., de La Reza, R. \& Bazzanella, B. 2007, MNRAS 377, 441]
\item[Patience, J., Ghez, A., Reid, I. N. \& Matthews, K. 2002, AJ 123, 1570]
\item[Prosser, C. 1992, AJ 103, 488]
\item[------- 1994, AJ 107, 1422]
\item[Prosser, C., Randich, S., \& Simon, T. 1998, AN 319, 215]
\item[Randich, S., Schmitt, J., Prosser, C. \& Stauffer, J. 1996, A\&A 305, 785]
\item[Rebull, L., Stapelfeldt, K., Werner, M. et al 2008, ApJ 681, 1484]
\item[Rhee, J., Song, I., Zuckerman, B. \& McElwain, M. 2007, ApJ 660, 1556]
\item[Robichon, N., Arenou, F., Mermilliod, J.-C. \& Turon, C. 1999, A\&A 345, 471]
\item[Rojo Arellano, E., Pena, J. \& Gonzalez, D. 1997, A\&AS 123, 25]
\item[Song, I., Zuckerman, B., Weinberger, A. \& Becklin, E. 2005, Nature 436, 363]
\item[Stauffer, J., Hartmann, L., Burnham, J. \& Jones, B. 1985, ApJ 289, 247]
\item[Stauffer, J., Hartmann, L., Fazio, G. et al 2007, ApJS 172, 663]
\item[Stauffer, J., Hartmann, L. \& Jones, B. 1989, ApJ 346, 160]
\item[Stauffer, J., Prosser, C., Giampapa, M., Soderblom, D. \& Simon, T. 1993, AJ 106, 229]
\item[Strassmeier, K. \& Fekel, F. 1990, A\&A 230,389]
\item[Su, K., Rieke, G., Stansberry, J. et al. 2006, ApJ 653, 675]
\item[Torres, C., Quast, G., Melo, C. \& Sterzik, M. 2008, in Young Nearby Loose Associations,]
Vol 5, ed. B. Reipurth, 757
\item[VandenBerg, D. \& Clem, J. 2003, AJ 126, 778]
\item[Weinberger, A., Becklin, E., Song, I. \& Zuckerman, B. 2011, ApJ 726, 72] 
\item[Wright, E., Eisenhardt, P., Mainzer, A. et al 2010, AJ 140, 1868]
\item[Zacharias, N., Finch, C., Girard, T. et al 2009, UCAC3 catalog]
\item[Zuckerman, B., Rhee, J., Song, I. \& Bessell, M. 2011, ApJ 732, 61]
\item[Zuckerman, B. \& Song, I. 2004, ARA\&A 42, 685]
 
\end{harvard}

\clearpage

\def\mc{\multicolumn}
\def\noIR{\tiny no IR source}
\def\str{\tiny Strange spec.}
\begin{landscape}
\begin{longtable}{lcccccccccccc}
\caption{High-Fidelity $\alpha$ Persei Cluster Members}\\
\hline
Star Name & RA&  Dec& pmRA& pmDec& Spectral& He& Pr?& Ma?& St?& Me?& Rob?& Ran? \\
& (J2000)& (J2000)& (mas/yr)&  (mas/yr) & type& & & & &  & & \\
\hline
\hline
\endfirsthead
\hline
\caption[]{(continued)}\\
Star Name & RA&  Dec& pmRA& pmDec& Spectral& He& Pr?& Ma?& St?& Me?& Rob?&  Ran?  \\
& (J2000)& (J2000)& (mas/yr)&  (mas/yr) & type& & & & &  & & \\
\hline
\hline
\hline
\endhead
\hline
\hline
\multicolumn{5}{r}{\small\sl continued on next page}\\
 \hline
 \endfoot
 \hline
 \endlastfoot
BD+48 851 & 03 07 49.8  &  +49 06 23  &  26.9$\pm$1.9  &  $-$23.6$\pm$1.9  & F6V& 12& y& y& & y&  & \\
TYC 3315 1159 1  &  03 11 16.8  &  +48 10 37  &  27.4$\pm$1.9  &  $-$28.0$\pm$1.8  &  F9V& 94& y& y& & & & \\
HD 19655    &  03 11 41.1  &  +48 03 15  &  31.0$\pm$1.4  &  $-$31.1$\pm$1.4  &  F2Vn& 104& y& y& & & & \\
HD 19624    &  03 11 42.9  &  +52 09 48  &  24.8$\pm$1.1  &  $-$25.1$\pm$1.2  &  B5& 145& y?& y& & & & \\
BD+49 868   &  03 11 50.0  &  +50 22 47  &  24.0$\pm$1.6  &  $-$23.4$\pm$1.6  & F5V& 135& y& y& & y & y & \\
HD 19767   &  03 12 43.3  &  +47 50 19  &  23.4$\pm$1.3  &  $-$25.3$\pm$1.3  &F0Vn& 151& y& y& &  & y & \\
HD 19805   &  03 13 05.2  &  +49 00 34  &  25.3$\pm$1.7  &  $-$25.6$\pm$1.8  &  B9.5V& 167& y& y&  & & y & \\
TYC 3319 446 1 &  03 13 07.4  &  +49 34 04  &  24.1$\pm$4.2  &  $-$30.2$\pm$3.8  &  G& 174& y& y& y?& y & &  \\
HD 19893 &  03 13 50.3  &  +49 34 08  &  26.1$\pm$1.2  &  $-$25.3$\pm$1.3  &  B9V& 212& y& y& & & y &\\
BD+48 871 &  03 15 23.6  &  +49 26 25  &  22.7$\pm$1.6  &  $-$21.6$\pm$1.6  & F7V& 270& y& y&  & y & y & \\
TYC 3319 9 1 &  03 15 58.9  &  +50 24 19  &  19.0$\pm$2.2  &  $-$17.5$\pm$2.0  &    F7& 299& y& y&  & y&  &\\
BD+49 889 &  03 16 23.2  &  +49 37 33  &  24.2$\pm$1.9  &  $-$21.8$\pm$1.9  & F5V& 309& y& y&  & y&  &\\
HD 20191 &  03 16 49.1&  +51 13 05  &  24.1$\pm$1.4  &  $-$23.9$\pm$1.5  & B9& 333& y& y&  &  & y &  \\
BD+49 892 &  03 16 59.4  &  +49 55 36  &  22.9$\pm$2.0  &  $-$21.8$\pm$2.0  & F7V& 334& y& y& y& y& &  \\
BD+48 876 &  03 17 20.5  &  +49 30 01  &  24.1$\pm$1.9  &  $-$26.1$\pm$1.9  & F7V& 338& y& y&  & y&  &\\
AP 119  &  03 17 31.4  &  +48 51 51  &  25.2$\pm$1.3  &  $-$23.5$\pm$2.2  & K2&  & y& & & & & \\
TYC 3319 306 1 &  03 17 36.9  &  +48 50 08  &  17.7$\pm$2.3  &  $-$19.6$\pm$2.2  &  G3& 350& y& y&  & y& & y \\
AP 121  &  03 17 42.1 &  +49 01 46  &  21.2$\pm$1.4  &  $-$21.1$\pm$1.3  & G5&  & y*y&  & y& ?& & \\
HD 20282  &  03 17 43.2  &  +50 21 40  &  31.2$\pm$1.6  &  $-$32.3$\pm$1.7  & A0& 357& & y& & & & \\
BD+49 896 &  03 18 01.7  &  +49 38 39  &  21.0$\pm$1.9  &  $-$23.0$\pm$1.9  &  F4V& 361& y& y&  & y& & \\
BD+49 897 &  03 18 05.2  &  +49 54 22  &  24.0$\pm$1.6  &  $-$24.0$\pm$1.6  & F6V& 365& y& y&  & & y & \\
HD 20344 &  03 18 23.9  &  +50 33 21  &  26.8$\pm$1.3  &  $-$26.9$\pm$1.5  &  A0& 379& y?& y& & & y & \\
V522 Per  &  03 18 27.4  &  +47 21 15  &  17.6$\pm$3.0  &  $-$26.9$\pm$2.7  &    G3& 373& y& y& & & & \\
29 Per    &  03 18 37.7  &  +50 13 20  &  22.7$\pm$1.0  &  $-$26.7$\pm$1.1  &  B3V& 383& y& y& & & y & \\
HD 20391 &  03 18 44.8  &  +49 46 12  &  21.5$\pm$1.8  &  $-$25.8$\pm$1.8  & A2V& 386& y& y&  & &  y& \\
TYC 3319 589 1 &  03 18 50.3  &  +49 43 52  &  20.1$\pm$2.2  &  $-$28.7$\pm$2.1  & G0& 389& y& y& & & &\\
V524 Per  &  03 18 59.3  &  +48 50 35  &  25.6$\pm$1.6  &  $-$23.7$\pm$2.0  &  K5& & y&  & & & & \\
31 Per      &  03 19 07.6  &  +50 05 42  &  21.9$\pm$1.2  &  $-$26.3$\pm$1.2  &  B5V& 401& y& y& & & & \\
HD 20475  &  03 19 42.2  &  +48 54 49  &  22.7$\pm$1.2  &  $-$27.1$\pm$1.3  & F2V& 421& y& y& & & y& y \\
HD 20487  &  03 19 47.2  &  +48 37 41  &  23.6$\pm$1.9  &  $-$24.4$\pm$2.0  & A0Vn& 423& y& y& & & y& \\
AP 126    &  03 19 57.3  &  +49 04 21  &  23.9$\pm$8.5  &  $-$21.6$\pm$8.5  &  M4&  & y&  & & & & \\
HD 20510  &  03 20 06.3  &  +50 58 07  &  23.0$\pm$0.9  &  $-$26.4$\pm$1.1  & B9V& 441& y?& y&  & & y& \\
HD 20537  &  03 20 23.7  &  +51 37 06  &  27.1$\pm$0.9  &  $-$26.5$\pm$1.0  &   B9& 450&  & y&  & & y& \\
TYC 3315 1781 1   &  03 20 39.3  &  +47 29 22  &  22.5$\pm$1.8  &  $-$18.8$\pm$1.7  & F8V& 453& y?& y& y& y& & y? \\
AP 97 & 03 20 41.9 & +48 24 38 &  23.9$\pm$1.4 & $-$25.3$\pm$1.3 & G6.5 &  & y$^a$ &  & y& y& & y \\
AP 131 &  03 20 56.5  &  +49 20 43  &  23.5$\pm$7.6  &  $-$26.9$\pm$7.6  &  M4&  &  y& & & &  & \\
V625 Per  & 03 21 06.5 & +48 26 13 &  18.7$\pm$1.8  &  $-$30.8$\pm$4.5  & G9&   &y$^a$ &  & y& y&  & y \\
AP 134  &  03 21 20.5  &  +47 53 15  &  23.7$\pm$7.9  &  $-$20.7$\pm$7.9  &  M3&  &  y&  & & & & \\
BD+47 808  &  03 21 30.2  &  +48 29 38  &  23.2$\pm$1.2  &  $-$24.1$\pm$1.2  &  F1IVn& 481&  y& y& & & y & \\
BD+48 892  &  03 21 40.2  &  +49 07 13  &  24.0$\pm$1.3  &  $-$26.6$\pm$1.4  & F3IV-V& 490& y& y&  & y & &\\
V459 Per  &  03 21 58.6  &  +49 12 53  &  23.5$\pm$1.3  &  $-$25.7$\pm$1.3  & F0IV& 501& ?& y&  & & &\\
V529 Per  &  03 22 06.8  &  +47 34 07  &  17.3$\pm$4.4  &  $-$28.3$\pm$4.0  &  K2&  & y&  &  &  & & y \\
V484 Per  &  03 22 21.9  &  +49 08 28  &  21.9$\pm$2.5  &  $-$25.0$\pm$2.3  &  G6V& 520& y& y& & & & y \\
HD 20714 &  03 22 26.3  &  +51 39 39  &  19.6$\pm$1.3  &  $-$20.1$\pm$1.4  &  A7Vn& 522& ?& y&  & & & \\
V575 Per  &  03 23 13.2  &  +49 12 48  &  20.6$\pm$1.3  &  $-$26.5$\pm$1.4  &  B5V& 557& y& y&  & &  y & \\
BD+47 815 &  03 23 40.3  &  +47 57 30  &  24.9$\pm$1.8  &  $-$28.4$\pm$1.8  & F3V& 577& ?& y& y?& & & y? \\
HD 20842 &  03 23 43.1  &  +51 46 13  &  23.2$\pm$1.4  &  $-$23.9$\pm$1.5  &  A0V& 575& y& y& & & & \\
HD 20863 &  03 23 47.3  &  +48 36 16  &  23.2$\pm$1.5  &  $-$27.0$\pm$1.5  &   B9V&  581& y&  y& & & y& y  \\
BD+ 49 914&  03 23 55.1 &  +50 18 24  &  23.1$\pm$1.6  &  $-$26.1$\pm$1.6  & F5V& 588& y& y& & & & \\
HD 20903  &  03 24 06.5  &  +46 17 19  &  25.4$\pm$1.1  &  $-$23.9$\pm$1.1  &     A2& & & y& & & & \\
AP 109 & 03 24 06.7  &  +49 24 52  &  15.4$\pm$8.4  &  $-$29.6$\pm$6.8  &  M3&  &  y$^a$&  & &  & & y\\
Melotte 20 601 & 03 24 17.1 & +49 39 00 & 21.5$\pm$1.5 &  $-$22.2$\pm$1.5  & G6 & 601& n&  &  y& y&  & \\
V461 Per &  03 24 19.2  &  +49 13 16  &  21.0$\pm$1.5  &  $-$24.8$\pm$1.5  & A8V& 606& y?& y& & & & y? \\
$\alpha$ Per &  03 24 19.4  &  +49 51 40  &  22.3$\pm$0.6  &  $-$26.1$\pm$0.7  &  F5Iab& 605& y& y& & & y& \\
AP 14  &  03 24 19.9  &  +48 47 20  &  22.8$\pm$1.2  &  $-$25.1$\pm$0.5  &   G4& & y&  & & y  & & y \\
V485 Per &  03 24 25.1  &  +48 48 21  &  15.1$\pm$6.5  &  $-$29.3$\pm$6.7  &  K5&  & & &  & & & y\\
BD+49 918 &  03 24 25.6  &  +50 19 34  &  22.5$\pm$1.4  &  $-$27.5$\pm$1.4  &  F0V& 609& y& y& & & & \\
HD 20931 &  03 24 30.0  &  +49 08 24  &  22.9$\pm$1.9  &  $-$25.6$\pm$2.0  &  A1V& 612& y& y& & & y& \\
BD+47 816&  03 24 47.1  &  +48 24 42  &  22.0$\pm$1.5  &  $-$26.3$\pm$1.5  & F4V& 621& y& y& & y& y&   \\
V531 Per  &  03 24 49.7  &  +48 52 18  &  23.2$\pm$3.0  &  $-$27.2$\pm$2.8  & G5& 622& y& y& y&  & & y \\
HD 20961 &  03 24 52.1  &  +47 54 54  &  23.7$\pm$1.6  &  $-$26.0$\pm$1.7  &  B9.5V& 625& y& y& & & & \\
BD+46 745 &  03 24 54.6  &  +47 24 54  &  22.1$\pm$1.4  &  $-$25.9$\pm$1.4  &   F4V& 632& y& y& & & y& \\
HD 20969 &  03 25 04.4  &  +49 47 43  &  22.0$\pm$1.3  &  $-$25.7$\pm$1.4  & A8V& 635& y& y& & & & \\
HD 20986 &  03 25 10.0  &  +49 15 06  &  22.0$\pm$1.3  &  $-$24.4$\pm$1.3  & A3Vn& 639& y& y& & & &\\
AP 25  &  03 25 16.2  &  +48 22 24  &  15.6$\pm$4.5  &  $-$25.1$\pm$4.0  &  K0& & y&  & & y& & y \\
HD 21005  &  03 25 20.7  &  +49 18 58  &  21.5$\pm$1.3  &  $-$25.1$\pm$1.3  & A5Vn& 651& y& y& & & & \\
TYC 3320 1768 1&  03 25 37.6  &  +50 19 18  &  19.8$\pm$1.5  &  $-$25.6$\pm$1.5  & F5V& 660& y?& y& y?& & &  \\
HD 21046 &  03 25 37.7  &  +47 01 14  &  22.6$\pm$1.0  &  $-$25.6$\pm$1.0  &  A7V& 665& y& y& & & & \\
V576 Per  &  03 25 57.4  &  +49 07 15  &  21.8$\pm$1.5  &  $-$28.2$\pm$1.5  &  B7V& 675& y& y& & & y& \\
V688 Per  &  03 26 04.2  &  +48 48 07 &  21.6$\pm$2.2  &  $-$22.7$\pm$2.2  &  F9V& 684& y& y& y& y& & y \\
HD 21091 &  03 26 10.8  &  +48 23 03  &  24.3$\pm$1.4  &  $-$28.7$\pm$1.5  &  B9.5V& 692& y?& y& & & y& \\
AP 38 &  03 26 19.3  &  +49 13 32  &  20.6$\pm$1.5  &  $-$28.1$\pm$3.0  &  G3& 696& y& & y& y& & y \\
V532 Per  &  03 26 22.2  &  +49 25 37  &  18.4$\pm$3.0  &  $-$19.0$\pm$2.7  & G2-3V& 699& y& y& & & & y \\
V628 Per  &  03 26 25.3  &  +48 20 07  &  20.5$\pm$1.7  &  $-$27.2$\pm$2.8  &  G5& & y&  & & y & & y \\
HD 21122 &  03 26 32.6  &  +47 15 59 &  24.5$\pm$1.1  &  $-$28.6$\pm$1.2  & A0& 710&  & y&  & & y &  \\
AP 158 &  03 26 33.7  &  +50 13 54  &  22.1$\pm$1.5  &  $-$25.4$\pm$0.9  & K0& & y&  & y& y& & \\
HD 21117  &  03 26 39.4  &  +50 50 47  &  24.9$\pm$1.2  &  $-$27.4$\pm$1.4  &  B8& 703& & y& & & y& \\
BD+47 825 &  03 26 39.2  &  +47 52 56  &  21.5$\pm$1.5  &  $-$26.0$\pm$1.5  & F2Vn& 721& ?& y& & & &  \\
TYC 3320 1715 1 &  03 26 40.7  &  +48 46 37  &  20.4$\pm$1.9  &  $-$25.8$\pm$1.9  & F4V& 715& y& y& & & & y\\
TYC 3320 818 1   &  03 26 43.9  &  +49 54 34  &  18.3$\pm$2.3  &  $-$23.4$\pm$2.3  &   G0V& 709& y& y& & & & y \\
AP 51 &  03 26 50.7  &  +48 47 31  &  20.0$\pm$2.2  &  $-$23.5$\pm$2.1  &  F7V& 727& y& y& & & & y\\
BD+48 916 &  03 27 03.2  &  +48 47 13  &  22.0$\pm$1.9  &  $-$25.4$\pm$1.9  &  F6V& 733& y?& y& & y & & y? \\
HD 21181 &  03 27 05.1  &  +48 12 20  &  23.9$\pm$1.3  &  $-$25.5$\pm$1.4  & BVn& 735& y& y& & & y&  \\
AP 161 &  03 27 18.9  &  +47 25 24  &  22.6$\pm$7.3  &  $-$19.9$\pm$7.0  & M3& & y&  & & & & \\
HD 21239  &  03 27 37.6  &  +48 16 23  &  23.4$\pm$1.6  &  $-$27.1$\pm$1.7  & A3Vn& 756& ?& y& & & &  \\
AP 58 &  03 27 37.8  &  +48 59 29 &  15.8$\pm$2.2  &  $-$26.1$\pm$2.1  & F9V& 750&  y& y& & & & y \\
HD 21238 &  03 27 38.9  &  +49 36 00  &  22.6$\pm$1.5  &  $-$25.3$\pm$1.5 & B9V& 747&  & y&  & & y& \\
V534 Per  &  03 27 51.0  &  +49 12 10  &  20.7$\pm$0.5  &  $-$24.9$\pm$1.4  & K2&  &  y& & & & & y \\
TYC 3320 2239 1  &  03 27 55.0  &  +49 45 37  &  18.3$\pm$2.3  &  $-$24.7$\pm$2.2  & F9V& 767& y& y& n?& y& &  \\
HD 21279  &  03 27 55.8  &  +47 44 09  &  23.9$\pm$1.7  &  $-$27.4$\pm$1.8  &  B8.5V& 775& y& y& & &  \\
HR 1034 &  03 28 03.1  &  +49 03 47  &  22.9$\pm$1.3  &  $-$26.0$\pm$1.3  &  B5V& 774& y& y& & & y&  \\
HD 21302 &  03 28 18.6  &  +49 57 10  &  22.5$\pm$1.7  &  $-$24.6$\pm$1.7  & A1Vn& 780& y& y& & &  & y \\
V488 Per & 03 28 18.7 & +48 39 48 & 15.7$\pm$6.6 &  $-$24.9$\pm$9.1  &  K0& & y&  &  & y& & y \\
BD+48 923  &  03 28 31.5  &  +48 56 27  &  22.2$\pm$1.9  &  $-$27.1$\pm$1.9  &  F4V& 799& y& y& & &  & y \\
BD+49 939  &  03 28 34.7  &  +50 16 01  &  22.6$\pm$1.4  &  $-$25.7$\pm$1.4  &  F3IV-V& 794& y?& y& & y&  &\\
HD 21345 &  03 28 38.0  &  +49 23 15  &  21.4$\pm$1.4  &  $-$25.0$\pm$1.4  &  A5Vn& 802& ?& y&  & & & \\
HR 1037  &  03 28 52.3  &  +49 50 54  &  21.7$\pm$1.6  &  $-$26.1$\pm$1.6  & B6Vn& 810& y& y& & & y&  \\
HD 21375 &  03 28 53.6  &  +49 04 13  &  24.6$\pm$1.9  &  $-$31.2$\pm$2.0  & A1V& 817& y& y& & & & y \\
Melotte 20 828& 03 28 59.6 & +48 14 08 & 19.2$\pm$1.3 & $-$26.0$\pm$1.7 & F8& 828& y&  & ?& y& & y \\
HD 21398 &  03 29 07.6  &  +48 18 10  &  22.5$\pm$1.5  &  $-$26.9$\pm$1.6  &  B9V& 831& y& y& & & & \\
TYC 3316 904 1 & 03 29 08.3 & +48 10 51 & 22.3$\pm$1.4  &  $-$27.6$\pm$1.0  & F6  & 833& y& &  & y& & \\
AP 176  &  03 29 19.0  &  +46 07 27  &  21.7$\pm$7.6  &  $-$27.6$\pm$7.4  & M4.2&  & y&  & & & &  \\
HD 21428 &  03 29 22.0  &  +49 30 32  &  19.4$\pm$2.2  &  $-$24.3$\pm$2.1  &  B3V& 835& y& y& & & & y\\
TYC 3320 1057 1 &  03 29 24.9  &  +48 57 45  &  24.1$\pm$2.0  &  $-$25.6$\pm$2.0  & F7V& 841& y& y& & & & y  \\
BD+47 837 &  03 29 26.2  &  +48 12 12  &  22.6$\pm$1.9  &  $-$28.1$\pm$1.9  &    F9V& 848& y& y& y& y&  & y\\
BD+48 931 &  03 29 46.9  &  +49 00 34  &  23.7$\pm$1.5  &  $-$28.0$\pm$1.5  &  F6V& 863& n& y&  & y&  & y\\
BD+47 839 &  03 29 46.9  &  +47 34 58  &  18.8$\pm$1.5  &  $-$27.7$\pm$1.6  &  F0& 876& y& y& & & & y \\
HD 21480 &  03 29 47.0  &  +49 09 13  &  27.5$\pm$1.5  &  $-$33.3$\pm$1.6  & A7V& 862& y& y& & & & \\
HD 21481 &  03 29 50.0  &  +47 58 37  &  22.9$\pm$1.1  &  $-$28.8$\pm$1.4  &    A0Vn&  875& n?& y& & & & y \\
HD 21479 &  03 29 51.8  &  +49 12 49  &  20.2$\pm$1.2  &  $-$27.2$\pm$1.4  &  A1IVn& 868& y& y& & &  &\\
HD 21527 &  03 30 19.3  &  +48 29 57  &  21.6$\pm$1.1  &  $-$28.3$\pm$1.1  &  A7IV& 885& y& y& & & y& y \\
V465 Per   &  03 30 34.0  &  +47 37 41  &  22.8$\pm$1.3  &  $-$24.5$\pm$1.4  &  A6Vn& 906& y& y& & & &  \\
HR 1051   &  03 30 36.9  &  +48 06 13  &  24.1$\pm$1.5  &  $-$23.5$\pm$1.6  &  B8V& 904& y& y& & &  y&  \\
Melotte 20 917 &  03 30 47.6  &  +47 53 22  &  22.3$\pm$1.3  &  $-$27.5$\pm$1.4  &   F4& 917& y&  & & y& & y  \\
AP 183  &  03 30 56.8  &  +50 00 51  &  19.6$\pm$7.5  &  $-$30.1$\pm$6.5  &  M3& & y& & & & &  \\
HD 21600  &  03 31 14.6  &  +49 42 22  &  21.7$\pm$1.4  &  $-$24.7$\pm$1.4  & A6Vn& 921&  y& y&  & & y&  \\
BD+48 937 &  03 31 29.0  &  +48 59 28  &  21.4$\pm$1.5  &  $-$26.9$\pm$1.4  & F9.5V& 935& y& y& & & & y \\
HD 21619 &  03 31 30.2 &  +49 54 07  &  21.7$\pm$1.4  &  $-$26.1$\pm$1.4  &  A6V& 931& y& y& & & y&  \\
HD 21641 &  03 31 33.1  &  +47 51 45  &  22.3$\pm$1.4  &  $-$25.9$\pm$1.5  &  B8.5V& 955& y& y& & & y&  \\
BD+49 957 &  03 31 44.2  &  +49 32 04  &  21.8$\pm$1.5  &  $-$25.5$\pm$1.6  & F3V& 944& y& y& & & & y \\
AP 189 &  03 31 44.9  &  +49 33 04  &  16.1$\pm$2.6  &  $-$21.1$\pm$1.5  & K3& & y?*y&  & & & & y  \\
HD 21672  &  03 31 53.9  &  +48 44 06  &  25.2$\pm$1.4  &  $-$29.0$\pm$1.5  &   B8V& 965& y& y& & & & y \\
TYC 3316 956 1 &  03 31 54.2  &  +48 31 38  &  22.6$\pm$1.7  &  $-$26.9$\pm$1.7  &  F8V& 968& y& y& y& y& & y  \\
BD+48 944  &  03 31 55.8  &  +48 35 02  &  25.3$\pm$1.4  &  $-$30.7$\pm$1.4  &   A4V& 970& y& y& & & & \\
BD+49 958  &  03 31 59.5  &  +49 52 10  &  24.0$\pm$1.3  &  $-$25.9$\pm$1.3  &  F1V& 958& n?& y& & & y&  \\
V396 Per  &  03 32 08.6  &  +48 01 24  &  23.2$\pm$1.4  &  $-$23.5$\pm$1.5  &  B8III& 985& y& y& & & y&  \\
V689 Per  &  03 32 10.2  &  +49 08 29  &  17.3$\pm$7.4  &  $-$25.3$\pm$7.3  &  G5V&  & y& & & & & y  \\
AP 196 &  03 32 19.3  &  +47 04 27  &  26.4$\pm$7.4  &  $-$31.7$\pm$7.4  &  K3& & y& & & y& &  \\
V537 Per &  03 32 30.7  &  +49 10 35  &  20.4$\pm$1.2  &  $-$27.2$\pm$1.1  &  G8&  & y& & & & & y \\
HD 232804 &  03 32 31.9  &  +51 29 22  &  21.0$\pm$1.4  &  $-$22.6$\pm$1.4  &     F5& 972& y& y& & & &  \\
HD 21855 &  03 33 22.2  &  +47 25 19  &  21.4$\pm$1.4  &  $-$27.6$\pm$1.4  &      A0& 1056& ?& y& & & y&  \\
BD+49 967   &  03 33 54.4  &  +50 17 48  &  21.5$\pm$1.4  &  $-$27.8$\pm$1.4  &   A6me& 1050& y?& y& & & y& \\
BD+50 784  &  03 33 58.9  &  +50 52 56  &  18.7$\pm$2.0  &  $-$26.0$\pm$1.9  &  F6& 1045& y& y& & & & \\
HD 21931 &  03 34 12.9  &  +48 37 03  &  22.5$\pm$1.5  &  $-$27.2$\pm$1.6  &   B9V& 1082& y& y& & & &  \\
BD+48 950  &  03 34 21.6  &  +48 39 36  &  22.2$\pm$1.6  &  $-$29.0$\pm$1.6  &  A2& 1084& y& y& & & &  \\
TYC 3325 239 1   &  03 35 05.0  &  +50 54 45  &  18.8$\pm$2.7  &  $-$22.3$\pm$2.5  &  G0:& 1086& y*y& y& y& y& &  \\
TYC 3321 1655 1 &  03 35 08.7  &  +49 44 39  &  19.1$\pm$2.7  &  $-$28.5$\pm$2.5  &  G4& 1101& y& y&  & y&  &  \\
HD 22136  &  03 35 58.5  &  +47 05 28  &  21.3$\pm$1.2  &  $-$24.5$\pm$1.2  &   B8V& 1153& y& y& & & y&  \\
V540 Per  &  03 36 22.0  &  +49 09 21  &  20.3$\pm$0.9  &  $-$24.9$\pm$1.0  & K0& 1143& y& & & & &  \\
$\psi$ Per &  03 36 29.4  &  +48 11 33  &  19.4$\pm$1.2  &  $-$28.9$\pm$1.2  &  B5Ve& 1164& n& y& & & y&  \\
TYC 3317 1064 1 &  03 36 31.8  &  +48 39 17  &  17.6$\pm$1.9  &  $-$27.7$\pm$1.9  &  F7& 1160& y& y& & & & \\
TYC 3321 187 1&  03 36 55.1  &  +48 49 43  &  20.8$\pm$1.9  &  $-$27.8$\pm$1.9  &  F7& 1180& y& y& & y& &  \\
TYC 3317 16 1 &  03 36 57.7  &  +48 44 46  &  19.2$\pm$2.6  &  $-$29.5$\pm$2.5  &  F7& 1185& y*y& y& & y& & \\
AP 229 &  03 37 27.5  &  +47 33 44  &  16.6$\pm$6.4  &  $-$30.5$\pm$6.4  &  K8& & y& & & & &  \\
HD 22401 &  03 38 15.6  &  +47 34 37  &  20.3$\pm$1.3  &  $-$27.4$\pm$1.4  &  A0V& 1259& y& y& & & y& \\
HD 22440 &  03 38 35.1  &  +48 35 37  &  19.4$\pm$1.6  &  $-$28.3$\pm$1.6  &     A2& 1260& y& y& & & y&  \\
TYC 3313 1551 1  &  03 38 51.0  &  +46 36 12  &  20.6$\pm$1.7  &  $-$25.9$\pm$1.7  &  G1& & & y& & & & \\
AP 248 &  03 41 04.8  &  +49 09 32  &  22.6$\pm$6.9  &  $-$29.2$\pm$6.8  &  M3& & y& & & & &  \\
HD 232823  &  03 41 40.9  &  +51 16 35  &  23.6$\pm$1.2  &  $-$30.1$\pm$1.3  &  F2& 1349& y?& y& & & & \\

\hline
\end{longtable}
\noindent Note $-$ Source names are from SIMBAD or Vizier. RA and Dec are from 2MASS. Most listed proper motions are from the Tycho-2 or UCAC3 catalog.  He = Heckmann et al 1956; Pr = Prosser 1992, if * is present, then also Prosser 1994; Ma = Makarov 2006; St = Stauffer et al 1993; Me = Mermilliod et al 2008; Rob = Robinchon et al 1999; Ran = Randich et al 1996. $^a$ = cluster member from Prosser et al 1998. 
In the cited papers, y = yes, member; y? = probably member; ? = membership
uncertain; n? = probably nonmember;  n = nonmember.  These evaluations are
taken directly from the cited papers and have not been reevaluated by the authors of the current study.
\end{landscape}

\clearpage

\def\mc{\multicolumn}
\def\noIR{\tiny no IR source}
\def\str{\tiny Strange spec.}
\begin{landscape}
\begin{longtable}{lcccccccccccc}
\caption{Possible $\alpha$ Persei Cluster Members}\\
\hline
Star Name & RA&  Dec& pmRA& pmDec& Spectral& He& Pr?& Ma?& St?& Me?& Rob?& Ran? \\
& (J2000)& (J2000)& (mas/yr)&  (mas/yr) & type& & & & &  & & \\
\hline
\hline
\endfirsthead
\hline
\caption[]{(continued)}\\
Star Name & RA&  Dec& pmRA& pmDec& Spectral& He& Pr?& Ma?& St?& Me?& Rob?& Ran? \\
& (J2000)& (J2000)& (mas/yr)&  (mas/yr) & type& & & & & & & \\
\hline
\hline
\hline
\endhead
\hline
\hline
\multicolumn{5}{r}{\small\sl continued on next page}\\
 \hline
 \endfoot
 \hline
 \endlastfoot
HD 17744  &  02 52 39.9 &  +48 48 59&  24.7$\pm$1.4  &  $-$25.5$\pm$1.5  & A0& &  & y& & & & \\
HD 18280  &   02 57 54.1 &  +48 52 43  &  23.2$\pm$1.6  &  $-$21.4$\pm$1.6  & A2&  & & y& & & & \\
BD+50 703  &   03 07 04.1  &  +51 17 40 &  17.0$\pm$1.5  &  $-$19.3$\pm$1.4  &   F2V& 7& ?& & & & &  \\
TYC 3318 1509 1 &  03 08 14.5  &  +49 50 21  &  23.5$\pm$3.3  &  $-$21.7$\pm$3.0  &  K0& & &y& & & & \\
BD+47 772  & 03 08 23.7  &  +47 36 16  &  19.4$\pm$1.7  &  $-$25.7$\pm$1.6  & F7& 13& & & & & & \\
HD 19458  &  03 09 24.2  &  +45 22 58  &  24.1$\pm$0.9  &  $-$23.1$\pm$1.0  & A0& & & y& & & & \\
HIP 14697  &  03 09 51.5  &  +48 28 18  &  23.6$\pm$1.9  &  $-$22.3$\pm$1.8  &  G3V& 56& y*n& y& n& & y & \\
V572 Per&  03 15 48.7  &  +50 57 21  &  33.2$\pm$1.6  &  $-$35.2$\pm$1.7  &  A0& 295& &y& & & & \\
HD 20135 &  03 16 01.5  &  +48 01 34  &  26.4$\pm$1.4  &  $-$18.4$\pm$1.4  &  A0p& 285& y& & & & &  \\
TYC 3319 915 1 &  03 17 24.0  &  +49 21 14  &  15.4$\pm$1.9  &  $-$24.7$\pm$1.9  & G0& 340& y?*y& & y& n & &  \\
TYC 3319 219 1 &  03 18 43.7  &  +50 23 11  &  26.6$\pm$1.3  &  $-$23.1$\pm$1.3  & F6IV-V& 387& y?&  & &  & & \\
TYC 3703 665 1  &  03 20 39.6  &  +52 43 35  &  28.1$\pm$2.6  &  $-$38.2$\pm$2.5  &  G3&  &  & y&  & & & \\
HD 232778  &  03 21 00.1  &  +50 59 35  &  19.4$\pm$1.3  &  $-$19.2$\pm$1.4  &  F8&  463& & & &  &  &  \\
HD 20701 &  03 22 03.5  &  +47 56 05  &  19.3$\pm$1.4  &  $-$23.3$\pm$1.4  &  A1V& 507& n& y&  & & &  \\
HD 21152 &  03 26 50.2  &  +47 54 58  &  20.5$\pm$1.7  &  $-$23.8$\pm$1.8  & B9V& 729& n?&  & & & &  \\
HD 232793&  03 27 14.1  &  +50 52 44  &  23.2$\pm$1.5  &  $-$26.0$\pm$1.5  & F5V& 732& ?&  y& & & &  \\
TYC 3316 677 1 &  03 27 36.7  &  +47 23 18  &  17.1$\pm$1.5  &  $-$24.0$\pm$1.5  & G1& 761& & & & &  & \\
AP 68  &  03 28 13.6  &  +49 13 13  &  20.9$\pm$6.9  &  $-$20.5$\pm$6.8  &  K3& &  & & & & & \\
AP 73  &  03 28 24.4  &  +48 53 46  &  15.2$\pm$2.3  &  $-$24.3$\pm$3.8  &  G3& & & & & & &  \\
TYC 3320 778 1  &  03 29 03.7  &  +50 21 12  &  22.6$\pm$2.5  &  $-$23.6$\pm$0.5  & G5& 818& & & & & &  \\
HD 21490 &  03 29 32.1  &  +43 13 50  &  22.8$\pm$1.2  &  $-$26.3$\pm$1.2  &  F2&  &  & y&  & & & \\
AP 85  &  03 30 22.0  &  +48 57 18  &  16.3$\pm$9.3  &  $-$25.2$\pm$6.3  &  K5& & & & & & &  \\
Melotte 20 1102 &  03 34 42.8  &  +47 53 14  &  19.6$\pm$2.3  &  $-$24.9$\pm$2.2  &  G1V& 1102& n?& y& & & & ? \\
BD+46 780  &  03 37 17.6  &  +47 20 54  &  27.9$\pm$1.1  &  $-$23.8$\pm$1.1  &  F3IV& 1218& y?& & & & &  \\
BD+51 756  &  03 39 02.9  &  +51 36 37  &  21.5$\pm$2.3  &  $-$34.1$\pm$2.2  & G2& 1234& y*y& y& & ?& & \\
TYC 3313 2091 1 &  03 41 05.7  &  +45 47 38  &  16.8$\pm$1.3  &  $-$25.2$\pm$1.2  & G1&  &  & y&  & & &  \\
HD 23219  &  03 45 27.5  &  +47 39 37  &  21.6$\pm$1.2  &  $-$27.2$\pm$1.4  & B9V& & & y& & & &  \\
HD 23287  &  03 45 53.2  &  +45 36 00  &  22.6$\pm$1.2  &  $-$28.5$\pm$1.3  &  A0& & & y& & & &  \\
HD 23255  &  03 45 54.7  &  +50 25 09  &  19.2$\pm$1.2  &  $-$26.2$\pm$1.2  &  A5& & & y& & & &  \\
HD 23690  &  03 49 01.0  &  +46 48 10  &  21.9$\pm$1.6  &  $-$26.4$\pm$1.7  &  A0& & & y& & & &  \\
HD 24260  &  03 53 39.5  &  +45 45 29  &  20.9$\pm$1.1  &  $-$26.0$\pm$1.2  &  A0& & & y& & & &  \\
HD 24980  &  04 00 20.5  &  +47 34 19  &  17.3$\pm$1.2  &  $-$25.9$\pm$1.2  &  A2& & & y& & & &  \\
HD 25109  &  04 01 50.0  &  +50 38 17  &  21.2$\pm$1.4  &  $-$29.6$\pm$1.5  &  A0& & & y& & & &  \\
\hline
\end{longtable}
\noindent Note $-$ Source names are from SIMBAD or Vizier. RA and Dec are from 2MASS. Proper motions are from the Tycho-2 or UCAC3 catalog.  He = Heckmann et al 1956; Pr = Prosser 1992, if * is present, then also Prosser 1994; Ma = Makarov 2006; St = Stauffer et al 1993; Me = Mermilliod et al 2008; Rob = Robinchon et al 1999; Ran = Randich et al 1996.
In the cited papers, y = yes, member; y? = probably member; ? = membership
uncertain; n? = probably nonmember;  n = nonmember.  These evaluations are
taken directly from the cited papers and have not been reevaluated by the authors of the current study.
\end{landscape}




\clearpage

\begin{landscape}
\begin{table}
\caption{Members of the $\alpha$ Persei Cluster with Excess Mid-Infrared Emission}
\begin{tabular}{@{}lccccccccccccccccc}
\br
Star& Spectral& & W1& (mJy)& & & W2& (mJy) & & & W3& (mJy) & & & W4& (mJy) & \\
& Type & Photo& Meas& Err& Ex& Photo& Meas& Err& Ex& Photo& Meas& Err& Ex& Photo& Meas& Err& Ex \\
\mr
HD 19624& B5& 570& 578& 19.7& -- & 318& 317& 6.1& -- & 54.1& 55.8& 0.88& -- & 15.1& 45.8& 1.5& 30.7 \\
HD 19893& B9& 466& 480& 14.6& -- & 261& 262& 4.6& -- & 44.4& 56.1& 0.83& 11.7& 12.5& 35.4& 1.6& 22.9 \\
V459 Per& F0& 144& 147& 3.3& -- & 81& 80& 1.5& -- & 13.8& 15.2& 0.27& 1.4& 3.9& 9.0& 0.87& 5.1 \\
HD 21091& B9.5& 339& 335& 8.7& -- & 190& 182& 3.5& -- & 32.2& 34.3& 0.54& 2.1?& 9.1& 18.9& 0.94&  9.8 \\
HD 21117 & B8& 348& 353& 9.1& -- & 195& 191& 3.5& -- & 33.3& 37.2& 0.55& 3.9 & 9.4& 17.8& 1.0& 8.4 \\
HD 21122 & A0& 252& 261& 6.0& -- & 142& 142& 2.9& -- & 24.1& 25.8& 0.4& 1.7?& 6.8& 13.7& 1.0& 6.9 \\
V 488 Per& K0& 21.9& 51.0& 1.1& 29.1& 9.3& 55.4& 1.1& 46.1& 2.1& 40.9& 0.6& 38.8& 0.57& 75.7& 2.2& 75.1  \\
HD 21375& A1& 441& 450& 14.1& -- & 247& 247& 4.6& -- & 42.1& 41.3& 0.6& -- & 11.9& 19.2& 1.1& 7.3 \\
HD 21480& A7& 234& 239& 5.5& -- & 132& 132& 2.6& -- & 22.4& 24.1& 0.4& 1.7?& 6.4& 14.7& 0.96& 8.3 \\
HD 21641& B8.5& 609& 609& 21.3& -- & 340& 343& 6.6& -- & 57.7& 70.7& 1.0& 13.0& 16.2& 39.9& 1.3& 23.7 \\
BD+50 784& F6& 111& 116& 2.5& -- & 60.4& 62.2& 1.2& -- & 10.8& 12.9& 0.24& 2.1& 3.1& 8.54& 0.9& 5.4 \\
\br
\end{tabular}
\end{table}
\noindent Note $-$ In addition to the above 11 stars, also HD 21855 may have weak excess emission in the W4 band.  Details may be found in Section 3.1.  See also discussion of HD 21641, V 488 Per, and HD 21375 in Section 3.1.  The four WISE wavebands W1, W2, W3 and W4 are 3.35, 4.60, 11.56 and 22.09 $\mu$m, respectively.
\end{landscape}

\clearpage

\begin{table}
\caption{Infrared Excess Fractions in Clusters/Associations}
\begin{tabular}{@{}lccc}
\br
Cluster/& age& 24 $\mu$m excess& Reference\\
Association& (Myr)& (\#) \ \ \ \ \ (\%)& \\
\mr
$\eta$ Cha& 6& 9/16   \ \ \ \ 56& Rebull et al (2008)\\
TW Hya Assoc. & 8& 7/23 \ \ \ \ 30& Rebull et al (2008) \\
UCL/LCC& 10& 10/35 \ \ \ 34& Rebull et al. (2008)\\
$\beta$ Pic MG& 12& 7/30 \ \ \ \ 23& Rebull et al (2008)\\
NGC 2547& 30& 16/38 \ \ \ 42& Gaspar et al (2009)\\
Tuc/Hor/Columba& 30& 27/62 \ \ \ 43.5& Zuckerman et al (2011) \\
IC 2391& 40& 6/26 \ \ \ \ 23& Rebull et al (2008) \& Gaspar et al (2009) \\
Argus Assoc. & 40& 5/8 \ \ \ \ \ 62.5& Zuckerman et al (2011) \\
NGC 2451& 60& 12/38 \ \ \ 32& Balog  et al (2009) \\
$\alpha$ Persei & 60& 8/56$^a$ \ \ \ \ 14$^a$&  this paper \\ 
AB Dor MG& 70& 12/47 \ \ \ 25.5& Zuckerman et al (2011) \\
Pleiades& 100& 10/73 \ \ \ \ 14&  Gorlova  et al (2006)\\
M47& 100& 8/63 \ \ \ \ 13& Rebull et al (2008)\\
Hyades& 650& 2/78 \ \ \ \ 2.5& Gaspar et al (2009)\\
Praesepe& 750& 1/135 \ \ \ \ 0.7& Gaspar et al (2009)\\ 
\br
\end{tabular}
\end{table}
\noindent Note $-$ $^a$For $\alpha$ Persei, these statistics pertain to stars in Table 1 of spectral types B and A and at a wavelength of 22 rather than 24 $\mu$m.  Should the excess IR emission at HD 21641 be due to an ionized outflowing wind rather than a dusty debris disk (see Section 3.1), then the entries for $\alpha$ Per in the third and fourth columns would read 7/56 and 12.5\%, respectively.  If the 12 A-type stars and B-star HD 23219 in Table 2 were to be added to the Table 1 stars, then entries in the third and fourth columns would read 8/69 and 11.6\%, respectively.   Three B-type stars from Table 1 with mid-IR excess emission ($\psi$ Per, 29 Per, and HR 1037, see Section 3) are excluded from all entries in Table 4 and in this caption.

\clearpage
\begin{figure}
\includegraphics[width=140mm]{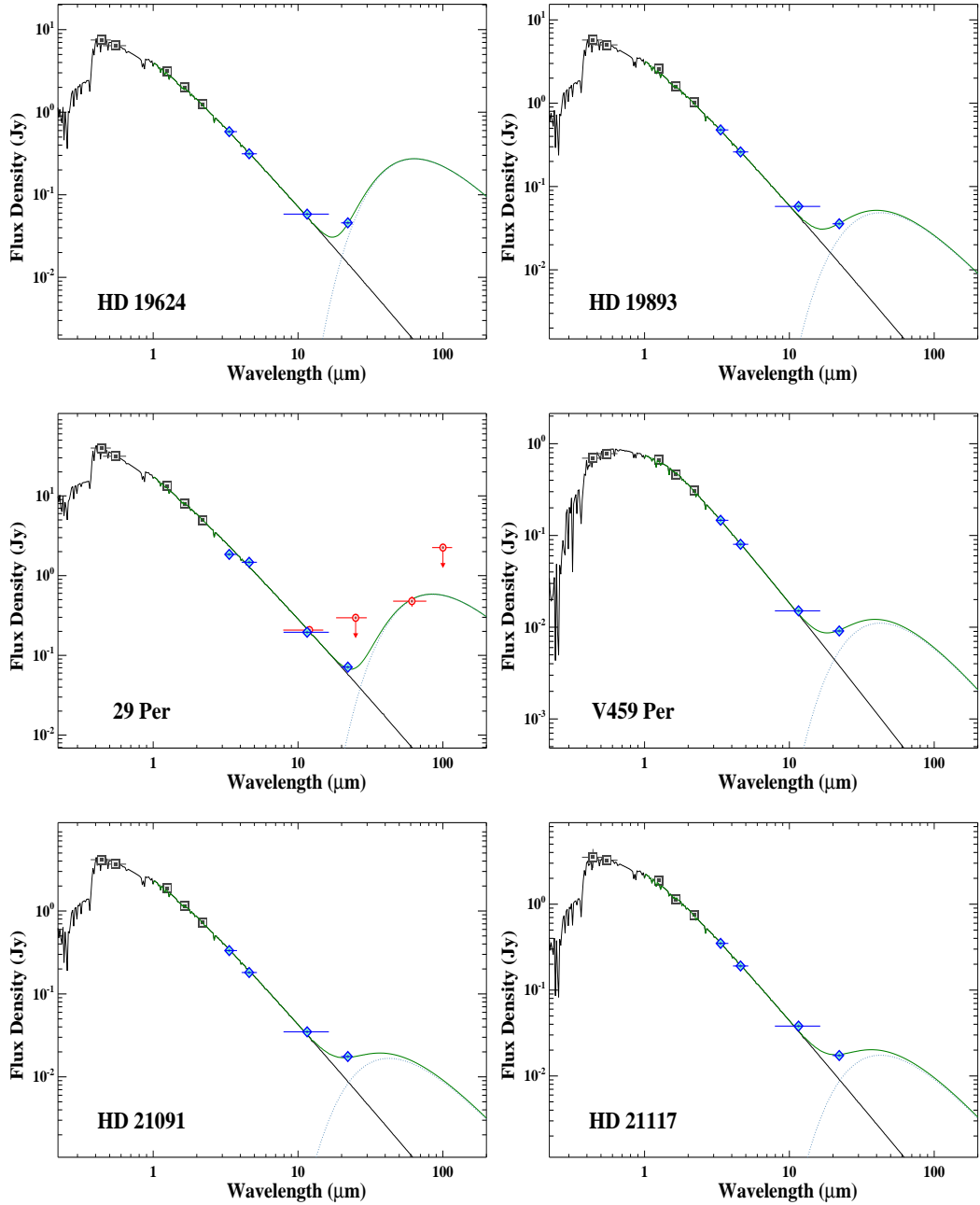}
\caption{\label{figure1} Spectral energy distributions (SED) for five $\alpha$ Persei cluster members in Table 3 with excess infrared emission and, in addition, the SED of 29 Per.  Near infrared JHK data points are from the 2MASS catalog.  The four diamonds are from WISE.  Circles between 12 and 100 $\mu$m for 29 Per are from IRAS.  The excess IR emission toward 29 Per is due to "cirrus" and not to an orbiting dusty debris disk (see Section 3.1).  Long wavelength blackbodies have been added to the various panels for illustrative purposes.  In all cases the blackbody temperature is 120 K, except for HD 19624 (80 K) and 29 Per (60 K).}
\end{figure}


\clearpage
\begin{figure}
\includegraphics[width=140mm]{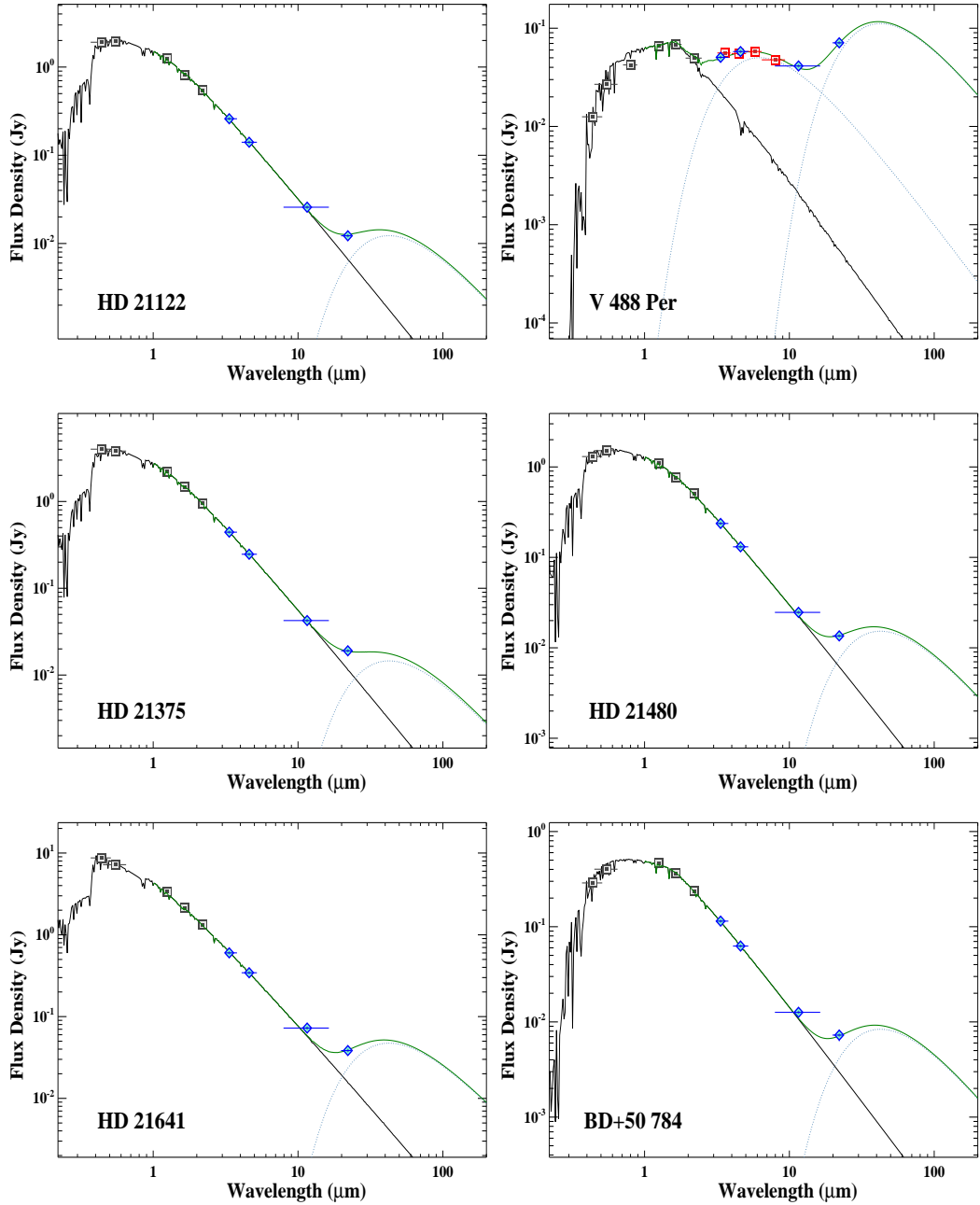}
\caption{\label{figure2} Same as for Figure 1.  The square data points for V488 Per between 3.5 and 8 $\mu$m are from the IRAC camera on Spitzer. The blackbodies indicated on the V488 Per panel have temperatures of 820 and 120 K.  The excess IR emission toward HD 21641 may not be due to a dusty debris disk (see Section 3.1).   Long wavelength blackbodies have been added to the various panels for illustrative purposes.  In all cases the blackbody temperature is 120 K.}
\end{figure}

\clearpage
\begin{figure}
\includegraphics[width=140mm]{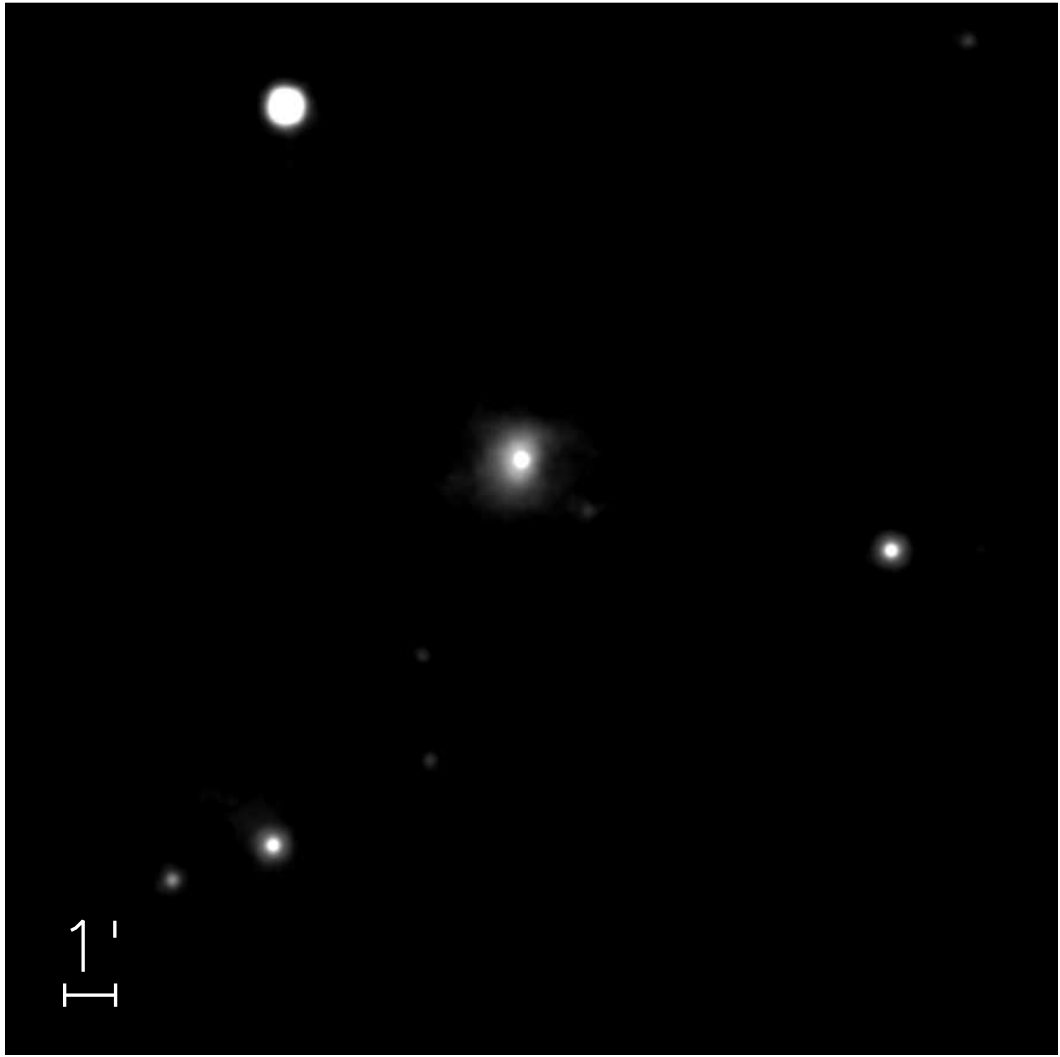}
\caption{\label{figure3} WISE 22 $\mu$m image of the area surrounding and including 29 Per (center
star in figure). North is up and East is left. The other sources are
unresolved and give an idea of the point spread function. The image is
presented with a logarithmic stretch.  29 Per is surrounded by an extended
nebulosity. The scale bar is 1 arcminute.}
\end{figure}

\clearpage
\begin{figure}
\includegraphics[width=140mm]{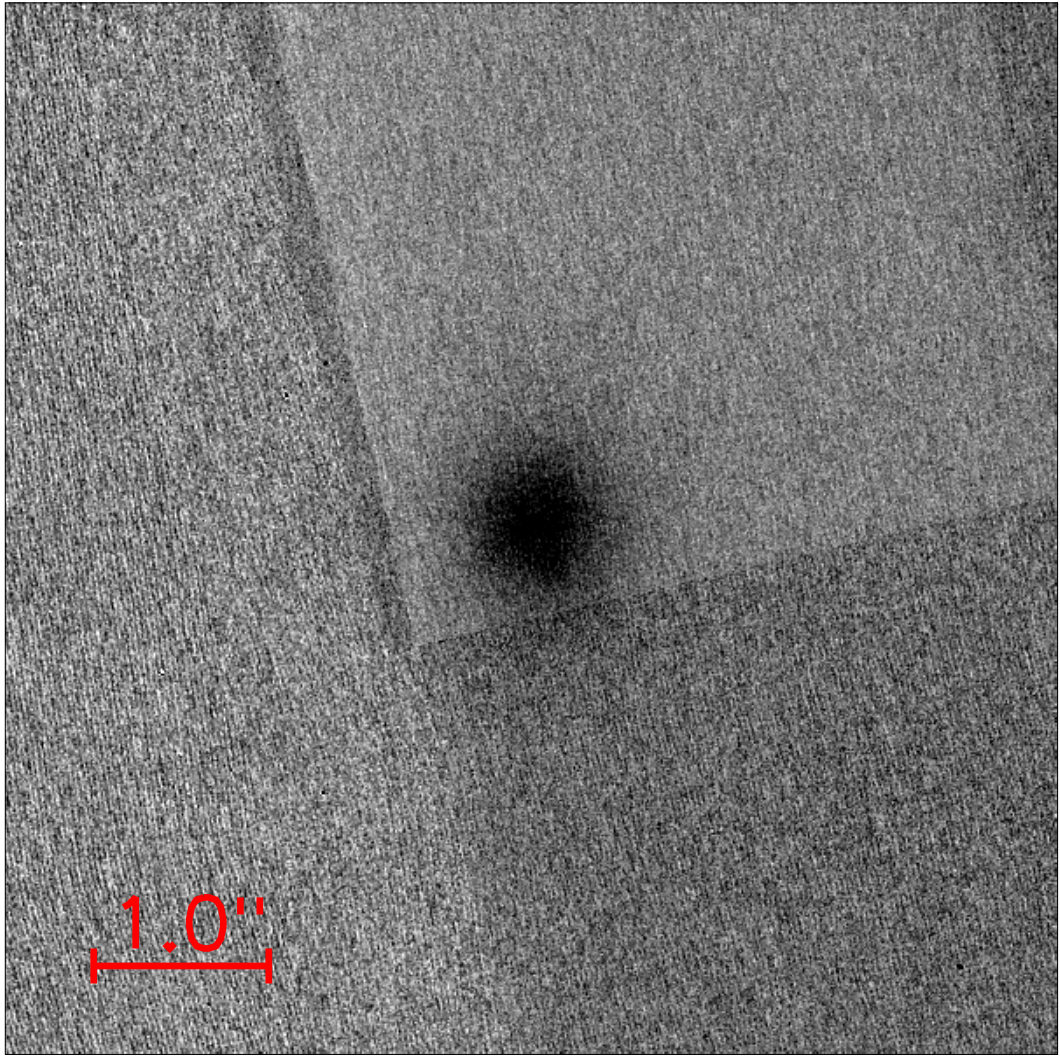}
\caption{\label{figure4} Seeing-limited image of V488 Per obtained with the Lp filter and NIRC2 camera on the Keck telescope on 31 March 2012 (UT).  The pixel scale is 0.01".  North is up, east to the left.  The displayed image is a combination of three 20 second exposures (0.2 sec times 100 coadds per exposure).  The images were obtained at an elevation of $\sim$1.6 air masses just before the telescope hit a hard limit, so there was not time to well-center V488 Per.  As a result, different noise levels are evident in various regions in the combined image which uses an inverted linear display.  Since the infrared excess emission at V488 Per at this wavelength is about twice as bright as the photosphere, if the excess were coming from a somewhat offset background object then this image would look like a double star.  The scalebar on the image indicates one arcsec; the WISE point spread function is about 6" FWHP at this wavelength.  Data courtesy of C. Beichman, C. Gelino, G. Mace, J. Aycock, and R. Campbell.}
\end{figure}

\clearpage
\begin{figure}
\includegraphics[width=140mm]{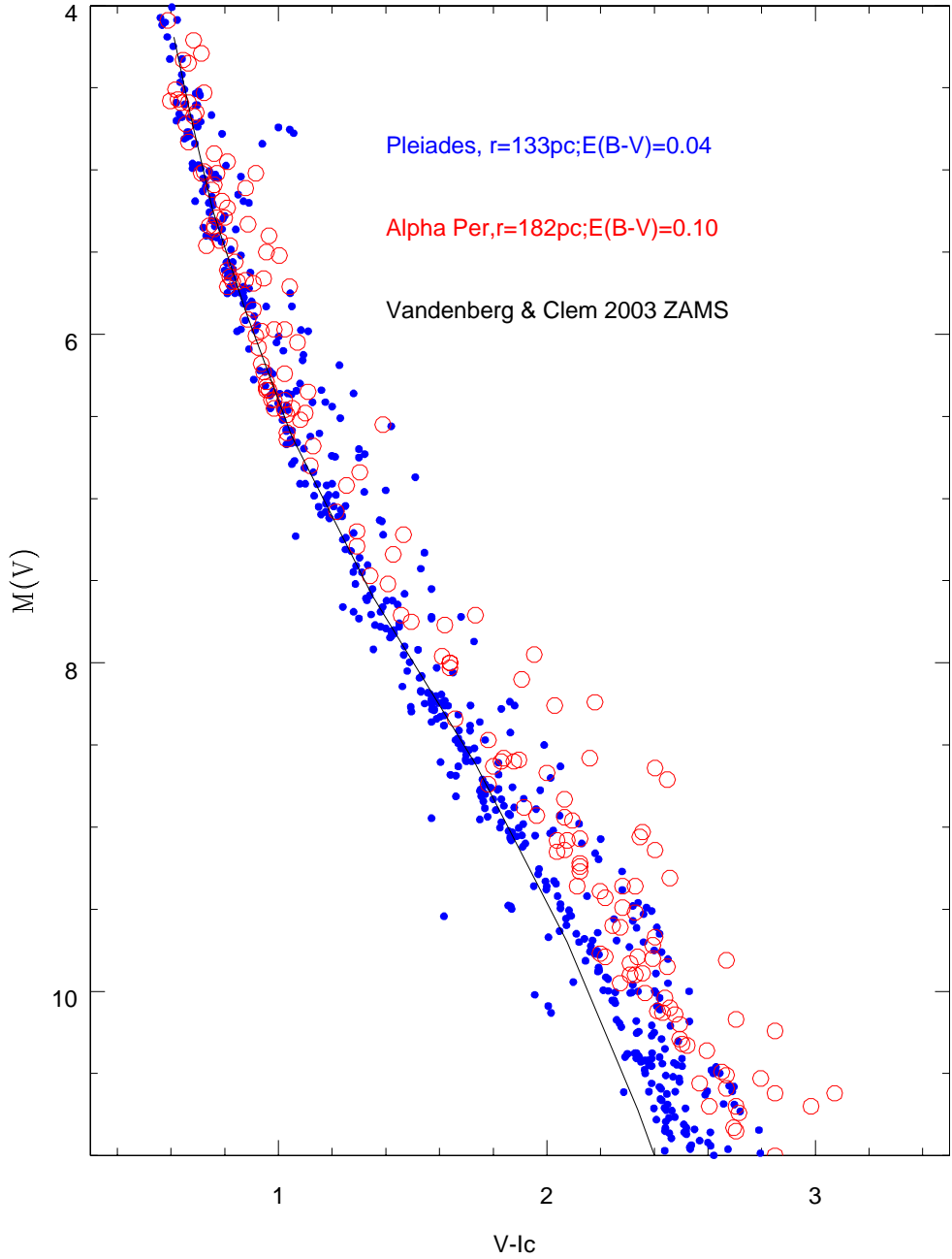}
\caption{\label{figure5} Comparison of color magnitude diagrams for the $\alpha$ Persei and Pleiades clusters (from B. Kamai et al 2012, in preparation).  The figure has not been cleaned of multiple star systems.  The data points for the Pleiades are from Stauffer et al (2007) and Kamai et al (2012, in preparation) and for $\alpha$ Per from Prosser (1992 and 1994) and Stauffer et al (1985 and 1989).  The relative placement of the late-type stars shows that $\alpha$ Per is younger than the Pleiades.}
\end{figure}

\end{document}